\documentclass[useAMS]{mn2e}
\usepackage{pslatex,graphicx}

\title[Circular Polarimetry and the Line of Sight to the BN Object]
{Circular Polarimetry and the Line of Sight to the BN Object}
\author[Aitken, Hough, Chrysostomou]
{D.K. Aitken,
J.H. Hough,
A. Chrysostomou \\
Centre for Astrophysics Research, Science \& Technology Research Institute, University of Hertfordshire, College
Lane, Hatfield, HERTS AL10 9AB, UK }

\newcommand{\gtsimeq}{\raisebox{-0.6ex}{$\,\stackrel
{\raisebox{-.2ex}{$\textstyle >$}}{\sim}\,$}}

\begin{document}
\label{firstpage}
\maketitle

\begin{abstract}

The 3.1\,$\mu$m absorption feature of water-ice has been observed
spectroscopically in many molecular clouds and,
when it has been observed spectropolarimetrically,
usually a corresponding polarization feature is seen.
Typically on these occasions, and particularly for the BN object,
a distinct position angle shift between the feature and 
continuum is seen, which indicates both a
fractionation of the icy 
material and a changing alignment direction along
the line of sight.

Here the dependence of circular polarimetry on fractionation
along the line of sight is investigated and it is shown
that the form of its spectrum, together with the sign
of the position angle shift,  indicates where
along the line of sight the icy material lies.
More specifically a coincidence  between the sign of the position angle
displacement in the ice feature, measured north through east,
and that of the circular polarization ice feature
means that the icy grains are overlaid by bare grains.
Some preliminary circular polarimetry of BN has this characteristic 
and a similar situation is found in the only two other cases for which relevant observations so far exist.

\end{abstract}


\section{Introduction}

In astronomy various materials may lie along the line of sight and their 
nature clarified by spectroscopy and spectropolarimetry; where
these materials may lie can only be inferred.
Circular polarimetry provides a means that can
directly probe relative position along the line of sight.

Molecular vibrations in the solid state give rise to features in the
near and mid-infrared which are characteristic of the chemistry of
the material.  Infrared spectroscopy has been used to
probe the chemical nature of interstellar dust, the solid component of the
interstellar medium (ISM).  In this way it has been found that 
the dust exists as small submicron sized grains
of which a ubiquitous component is a silicate-like material, but
other constituents without an infrared signature such as carbon
must be present, 
and the existence of various ices and refractory materials
depend on the local environment: for example ice features
are only observed within dense clouds.

Because the grains are in general non-spherical and are often aligned and
oriented by the ambient magnetic field the
extinction will differ for
{\bf E} vectors parallel and orthogonal to this projected direction
(dichroism) and the
the radiation traversing the medium becomes polarised.
A phase change between these orthogonal directions is also
introduced (birefringence)  and circular polarization will be produced
if the alignment direction changes along the line of sight (Serkowski 1962).
Apart from very small effects the position angle remains independent
of wavelength except if the dust composition also changes along the
line of sight, in which case the position angle of
the different molecular species will be different. Thus a
changing position angle at the characteristic wavelengths of different
dust species indicates both fractionation and changing alignment
direction.
Such effects are often observed when linear polarization studies have
been made of ice features towards sources in dense clouds
(eg Hough etal 1989, 1996; Holloway etal 2002),
and  only polarimetry can demonstrate this line
of sight structure so directly.
In the near infrared
circular polarization has been obseved in the direction of the Becklin
Neugebauer (BN) object in Orion by Serkowski and Rieke
(1973), and from several molecular cloud sources by Lonsdale etal (1980)
and Dyck and Lonsdale (1980) and a twist of alignment direction must
exist to yield the observed circular polarization.
Hough etal (1996) observed a position angle change of several degrees in the
linear polarization of the ice feature in BN and this additionally requires that
the icy material is fractionated along the line of sight.  Neither
the intensity spectrum nor the polarization spectrum are sensitive to
the details of the fractionation other than the existence
of an alignment twist and 
the sign of the angular displacement between the components.
Conversely circular polarization does depend on the sign of twist and on the
ordering of materials along the line of sight;  in this sense it is not
commutative, and the purpose of this paper is to show how this can be
turned to practical use.

\section{Circular polarization from changing grain alignment}

\begin{figure*}
\includegraphics*{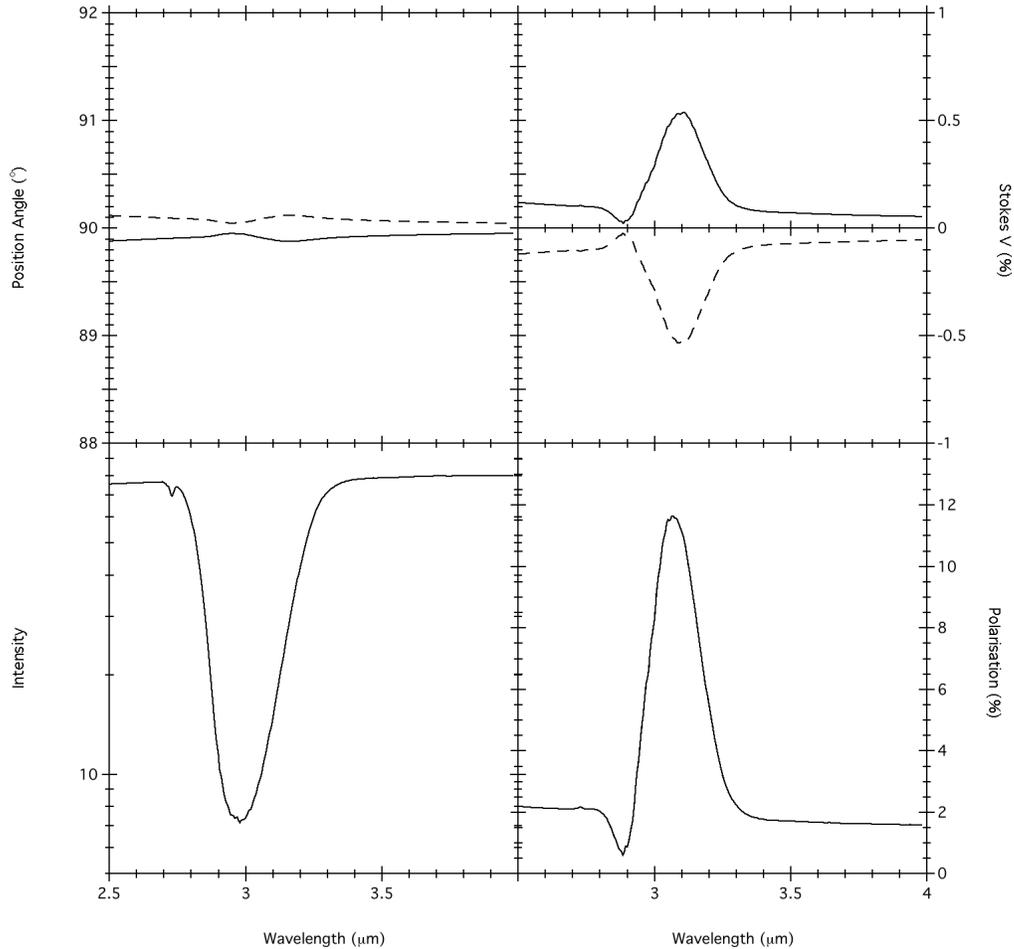}
\caption{Intensity, linear polarization, position angle and circular
polarization produced by two identical regions of H$_2$O mantles on silicate
grains with grain alignments at 100$^{\circ}$ and 80$^{\circ}$. Solid lines denote the 100$^{\circ}$ region preceeding the 80$^{\circ}$ region and dotted lines the inverse. Extinction (hence the intensity) and linear polarization are unaffected by reversal while circular polarization inverts and the position angle departs very slightly from the mean of the alignment angles as it is displaced towards the last alignment direction. The position angle is weakly wavelength dependent. The grains are of `astrosilicate' cores with mantles of pure water-ice.}
\label{fig1}
\end{figure*}

\begin{figure*}
\includegraphics*{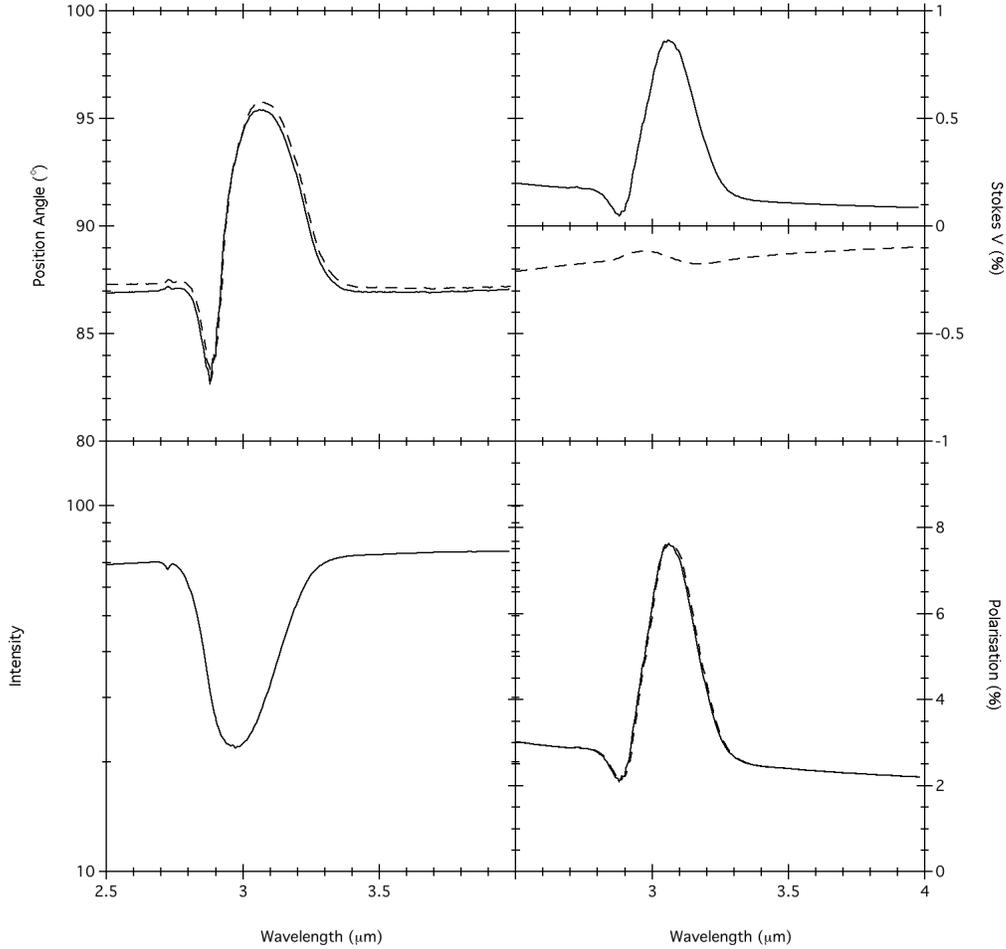}
\caption{As for Fig 1 except that the icy mantles have been removed from
the second region. The ice feature vanishes from circular polarization
on reversal and the  position angle spectrum is barely changed.}
\label{fig2}
\end{figure*}

\begin{figure*}
\includegraphics*{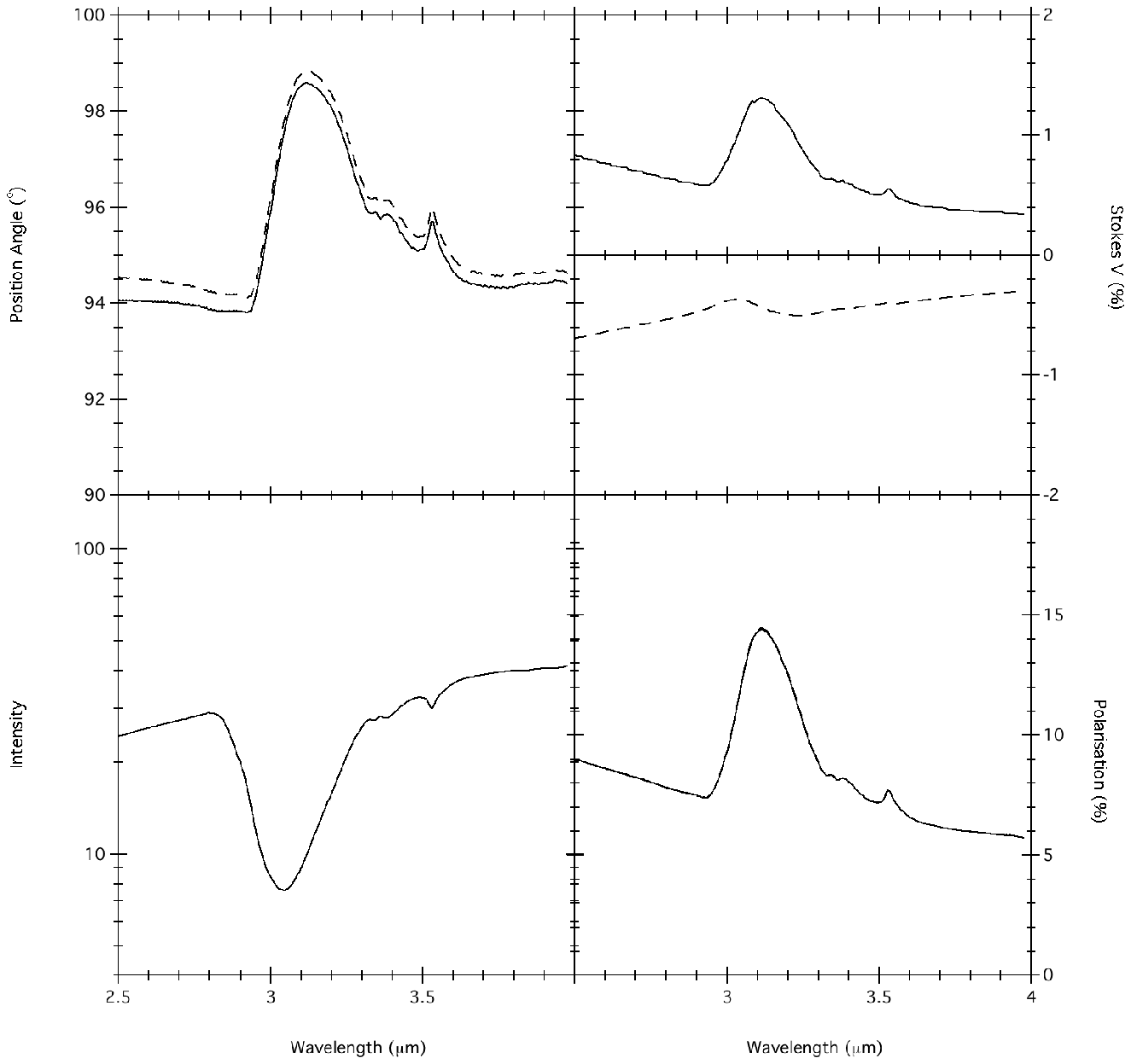}
\caption{Observed circular and linear polarizations in the region
of the ice feature for model AC13 in Table 1.}
\label{fig3}
\end{figure*}

Circular polarization can arise in a number of ways but
in the interstellar medium always as a secondary process:
through (a) scattering of polarized light
by dust grains, (b) scattering of unpolarized light by aligned
non spherical grains, (c) by passage of linearly polarised light through a
medium of aligned nonspherical grains.  In the latter case the medium is
birefringent through the presence of aligned grains and circular
polarization is produced so long as the alignment direction is neither
parallel nor orthogonal to the linear polarization.
In the nearer infrared the former two processes are important and can yield
high percentatages of circular polarization, but it
is the latter process that is of interest here as
the sign and spectral form of the circular polarization depend
on the twist between the regions and their distribution along the line
of sight.

Polarization produced by a medium with changing grain alignment
is discussed by Martin (1974, 1978), who 
treats analytically the case of uniform slabs of differing polarization
properties at varying angles to each other, and also that of a continuous
medium with uniform twist.
Much of what is presented here was inspired by and
derives from Martin's 1974 paper
but the analysis concentrates
more explicitly on what can be learned from the 
combination of linear and circular polarimetry.
In the two  slab case Martin (1974) points
out that the circular polarization, $V/I$, depends on the linear
dichroism of the first slab, the linear birefringence of the second slab,
and the angle between the two slabs, so that it is sensitive to
the sequence of material changes along the line of sight.
In his treatment Martin neglects
some minor terms to make the integration more tractable and, because
the treatment is analytic, he derives useful and instructive relationships 
governing both the amount and sign of the circular polarization and the
amount of position angle change in the linear polarization.
In this study a numerical approach  is adopted which is more convenient for
the display of the wavelength dependence of specific models of ice mantled
grains and can use the full polarization transfer.

Following Martin (1974) two slab and uniformly twisted models are considered.
For the two slab case
the two regions along the line of sight each contain a
uniform grain population in respect of shape, chemistry and alignment.
These properties can differ between the two regions as can their extinction.
In the single twisted region the change in grain properties is introduced
at a discrete twist angle.
The initial radiation is unpolarized and independent of
wavelength, a simplification that does not affect the polarization. 
Grains are taken as oblate with cross sections
for absorption and phase determined from the real and imaginary parts of the 
electric polarizability in the Rayleigh limit (eg Draine and Lee 1984) and
grain mantles are assumed confocal.
The polarization cross sections for oblate grains,
$C_{pol}= C_{abs}^\perp -C_{abs}^\parallel$, where
$C_{abs}^\perp$ and $C_{abs}^\parallel$ are the
absorption cross sections
for {\bf E} vectors perpendicular and parallel to the symmetry axes,
determine the dichroism of the medium along the line of sight
through $C_{pol}\,R\,$cos$^2\delta$,
 where $\delta$ is the angle between alignment direction
and the plane of the sky.
Here $R= \frac{3}{2}(<$cos$^2\beta>-\frac{1}{3})$ is the Rayleigh reduction
factor  for spinning grains precessing with angle $\beta$ about the
alignment direction (Greenberg 1968).  Thus $C_{pol}\,R\,$cos$^2\delta$
is the effective  cross section for polarized intensity
in the plane of the sky.
In a similar way $C_{crc}= C_{pha}^\perp -C_{pha}^\parallel$,
where the $C_{pha}$'s are the respective cross sections  for phase advance,
determine the birefringence of the medium
through $C_{crc}\,R\,$cos$^2\delta$. 
The extinction cross section is
$$C_{abs}=  \frac{2C_{abs}^\perp +C_{abs}^\parallel}{3}-
    \frac{RC_{pol}}{6}(3\rm{cos}^2\delta-2)$$
and the second term can usually be ignored. Expressions for
$C_{abs}^\perp$, $C_{abs}^\parallel$, $C_{pha}^\perp$ and  $C_{pha}^\parallel$ for oblate grains follow  Draine and Lee (1984).

The polarization transfer equations (see Appendix)
are iterated through each slab in turn and the two slabs are then reversed to determine effects due to ordering along the line of sight.  

Martin (1974) gives a number of relations between the linear and
circular polarization and the position angle
which can be applied in the case of two slabs.
Denoting by $\theta_{1}$ and $\theta_{2}$ the alignment direction
in the first and second slabs and by $r$ the ratio $p_{2}/p_{1}$ of
the linear polarizations produced in the second to first slabs then the
depolarization is given by $D=\sqrt{(1 + 2r\, \textrm{cos}(2\phi) +r^2)}/(1+r)$ (Martin 1974, equation 24), where $\phi=\theta_{2}-\theta_{1}$,
and the position angle, $\theta$, of the resultant is given by (Martin 1974,
equation 27)
tan\,$2(\theta-\theta_{1}) =r\,$sin$\,2\phi/(1+r\,$cos$\,2\phi)$. So long as the grains in the two slabs have the same
chemistry (and shape), $r=p_2/p_1$ is independent of wavelength
and so is the position angle, $\theta$.  If the grains in one of the slabs
displays extra polarization at some wavelength which changes $r$ to $r_f$
then there will be a position angle change
$$
\textrm{tan}\,2\Delta\theta = (r_f - r_c)\,\textrm{sin}\,2\phi/(1+r_f r_c +(r_f+r_c) \textrm{cos}\,2\phi),
$$
where $r_c$ is the polarization ratio at other wavelengths.  If the
slabs are interchanged the linear polarization and position angle
are unchanged ($\phi$ changes sign, $r_f$ and $r_c$ invert) in this
approximation.  Only the circular polarizaton changes, both in sign
and character.

Approximations in these relations compared to
a full polarization transfer depend on the dielectic functions
involved but are usually negligible
so long as the polarizations are small.
As an example  in Fig 1 both slabs 
contain identical ``astrosilicate" (Draine and Lee 1984, Draine 1985)
cores with water-ice mantles (Hudgins etal 1993)
at an angle of 20$^{\circ}$. Continuous
lines are for a 100$^{\circ}$ aligned slab preceeding an 80$^{\circ}$
slab and the broken line is the reversed situation.
There is no position angle change in the ice feature but a very small change
of $\sim 0.1^{\circ}$ displaced from the
mean towards the second slab in each case.  The circular polarization 
inverts and has a spectral form similar to the linear polarization whose
change is undetectable in the figure. 

As a further example we take the
slabs to contain grains with an ``astrosilicate" core with and
without a water-ice mantle;
Fig 2 shows that the linear polarization and
the position angle are independent of the order of the slabs
to a very close approximation,
while the circular polarization displays the
ice feature only when the icy
slab precedes the other along the line of sight.  After interchanging
just a reversed sign continuum remains with little evidence for structure,
and what structure remains is spectrally different
from either the scaled
down unreversed circular or linear polarization.
The small changes seen in the position angle
are due to the full polarization transfer;  very small changes to the linear
polarization are not discernable.

More generally, following Martin (1974),
for dissimilar slabs the circular polarization
is dominated by the spectral properties of the first slab, and its sign
by the handedness of the alignment twist; 
positive sign indicates an
alignment angle increasing (north through east)
with increasing depth, or equivalently a
clockwise twist in the direction towards the observer.
If the position angle shift of a component has he same sign as the 
circular polarization that component must be further down the line of sight (than what 
produces the continuum) and vice versa.

\section{A model of the line of sight to BN}

In Fig 2 the pure H$_{2}$O ice feature is narrower than the observed feature
and the background continuum level is not well reproduced by ``astrosilicate"
alone.
The extinction declines rapidly from its peak to $3.3\mu$m and then shows
a slower decrease in an extended wing to 3.6$\mu$m and this is reproduced
in the linear polarization with the wing being slightly more prominent.
Such structure, different from pure H$_{2}$O ice, 
has been observed in the spectrum of many young stellar objects
(Smith, Sellgren and Tokunaga 1989) and
attributed to the presence of other ices such as NH$_{3}$ and/or CH$_{3}$OH.
There is considerable variation in the extinction profile of the young
stellar objects observed by Smith, Sellgren and Tokunaga and
BN appears to be an extreme case with a pronounced narrow additional
absorption feature at 3.08$\mu$m, which was interpreted as requiring
a range of temperatures of H$_{2}$O ices up to 150K.

Modelling of the material along the line of sight to BN which includes
polarimetric observations has been done by Lee and Draine (1985).
As a starting point they used the MRN graphite-silicate model 
(Mathis, Rumpl, and Nordsieck 1977) using
bare oblate grains of `astronomical silicate' and graphite, and these
grain cores mantled with an H$_{2}$O:NH$_3$ mixture and
obtained a satisfactory match to the then existing linear polarimetry
around the ice feature and in the mid infrared
(Capps, Gillett and Knacke 1978, Capps 1976).
To reproduce the circular
polarization observations of Lonsdale {\em et al} (1980) at 2.2$\mu$m
and Serkowski and Rieke (1973) at 3.45$\mu$m they used a twist
angle of $\phi \simeq 25^{\circ}$ confined to the region containing the
core/mantled grains, and this predicted circular polarization close to 1\%
which indicated the presence of an ice feature.

Hough etal (1996) presented high quality spectropolarimetry of the
ice feature which revealed a distinct position angle shift  between
the feature and continuum.  
Because of this evidence of fractionation
of the ice along the line of sight they used a two slab model in which
bare and mantled grains were in separate regions with an angle of
10-25$^{\circ}$ between them, to produce the observed 4$^{\circ}$
position angle shift.

Modelling has also been presented by Holloway (2003)
who produced greatly improved fits in the
mid-infrared by using amorphous olivine (Dorschner etal 1995) in place
of `astrosilicate' and amorphous carbon (Zubko etal 1996)
in preference to graphite.  There are some small discrepancies
between the polarization properties of BN reported by Hough etal (1996)
and Holloway etal (2002) in the wavelength region around and short
of 3.0$\mu$m.
These differences concern the detailed shape of the polarization, and not
the levels of feature and continuum polarization and do not affect the
way the present modelling is performed.  Nevertheless it is important that
the spectropolarimetric observations of BN, and other objects, are repeated.

Following Hough etal 1996 we use the dielectric function of the ``strong" ice
mixture of Hudgins etal (1993), containing H$_{2}$O,
CH$_{3}$OH, CO and NH$_{3}$ in the ratio 100:50:1:1, but here use the
temperature of 120$^{\circ}$K as mantles on silicate and carbon cores,
together with bare grains of silicate and carbon.
Bare grains are taken as oblate with the mantles confocal on identical
cores, and cross-sections found in the Rayleigh limit using equations
in Draine and Lee (1984).
Draine's (1985) ``astrosilicate"  dielectric function
is used but the dielectric function for crystalline graphite depends on
the direction of its c-axis with respect to the radiation {\bf E} vector;
$\epsilon_{\parallel}$ and $\epsilon_{\perp}$
represent {\bf E} vectors parallel and perpendicular to this axis, and
are very different functions of $\lambda$.
An approximation is made (Draine and Lee 1984) that $1/3$ of the grains
have the {\em isotropic} dielectric function $\epsilon_{\parallel}$
and $2/3$ have the {\em isotropic} dielectric function $\epsilon_{\perp}$.
The corresponding cross sections are computed and their sum taken
as representing a random distribution of graphite c-axes within the grain;
bare and core graphite grains are both treated in this way.
Partly because of the approximations made for graphite,
cross sections based on an amorphous carbon dielectric function
from Zubko etal (1996)  are also used and referred to as ACAR.

Recently Robberto etal (2005) have placed the extinction to the BN object 
at $\tau_{9.8\mu}=1.37$ from a
fit to filter photometry through the silicate feature.  However they ignore
the presence of underlying silicate emission (Gillett etal 1975;
Aitken etal 1981) indicated by single aperture spectroscopy and the fit then
indicates a much larger extinction of $\tau_{10\mu}=3.3$, which is the
value for the $\tau_{10\mu}$ used here.
An additional and independent factor confirming this larger value is the
observed absorptive polarization of 12.5\% at 10$\mu$m
(Aitken, Smith, Roche, 1989),
making its specific polarization $p/\tau$~=~.125/3.3~=~.038,
the largest for galactic sources out of a sample of 30
(Smith etal 2000).
The constraint on the amount
of the ``strong" ice mixture is provided by the depth of the ice
extinction  $\tau_{3.1\mu}=1.6$.
The ratio of carbon/silicate has been taken 
 as 0.7 by number, which is between the MRN
value of .87 and that used by Lee and Draine, of .45 in their case A.

Further observational constraints are provided by the linear polarizations
of 12.5\% at 10$\mu$m, 16\%
at the peak of the ice feature at 3.1$\mu$m and its
continuum of 8.5\% at 3.6$\mu$m (Hough etal 1996).
At this stage no contraints are derived from the circular polarization
or the position angle of the linear polarization.
To restrict the number of free parameters bare grains
and grain cores of silicate and carbon are taken to be the same size and
oblate and the ratio of mantled to bare grains is taken as
the same for silicate as for graphite/carbon.

The above constraints and assumptions together with the cross sections
of the grains for extinction define the column density of the bare and the
mantled grains.
The number ratio  bare/mantled grains is then 2.7 if the
volume ratio mantle/core, 
$v_{m}/v_{c}$=1, and
4.55 if $v_{m}/v_{c}$=1.5, implying a correspondingly
larger column depth of bare grains than mantled.
Linear polarizations and polarization cross-sections  are then enough to
determine three of the polarization reduction factors   
in terms of the fourth and this restricts the
physically significant  reduction factors to a continuum range 
shown sampled in Table 1, for $b/a=2.0$ ($a$ is the symmetry axis and 
$v_{m}/v_{c}$=1:
$R_{sm}$ for ice mantles on silicates, $R_{gm}$ for ice on graphite/carbon,
$R_{s}$ for bare silicates, and $R_{g}$ for bare carbon.  
Estimated in this way the the reduction factors contain the cos$^2\delta$
term due to an alignment angle, $\delta$, out of the plane of the sky,
and the effect of any fluctuations along the line of sight,
so that the tabled reduction factors will represent lower limits.
Grains with  $v_{m}/v_{c}=1.0$ and $v_{m}/v_{c}=1.5$ for
$b/a=1.5$ were also considered but are not presented here as they
required seemingly implausibly large 
reduction factors to reproduce the observed linear polarizations.

This method does not attempt to fit properties such as feature width, shape
or precise peak wavelength which may depend on many minor constituents
and a range of physical conditions.
The main intention here is to use the major ingredients and find
a range of reduction factors which
reproduce the observed levels of extinction and linear polarization
in the feature and continuum and investigate how these are constrained 
by the observed position angle shift and what circular polarization
is predicted by different sequences.

\begin{tabular}{ccccc}
\multicolumn{5}{c}{\bf Table 1}\\
\hline
\multicolumn{5}{c}{Polarization reduction factors} \\
\multicolumn{5}{c}{     for b/a=2.0 and $v_{m}/v_{c}$=1.0 }\\
\hline
label &  \multicolumn{4}{c} {reduction factors } \\
  & $R_{sm}$ & $R_{gm}$ & $R_{s}$ & $R_{g}$ \\
  \hline
 \multicolumn{5}{c} {amorphous carbon (ACAR)}\\

\hline
AC11 &  .129&.269&.083&0\\
AC12&   .166&.214&.069&.025\\
AC13&   .189&.180&.060& .04\\
AC14&   .204&.158&.055&.05\\
AC15&   .241&.102&.041 & .075\\
AC16&   .279&.047&.027&  .1\\
AC17&   .310&.001&.015&  .121\\

\hline
  \multicolumn{5}{c} {graphite}\\

\hline
GR11 &  0&.275&.119&.0385\\
GR12&   .0194&.256&.112&.05\\
GR13&   .065&.211&.093& .075\\
GR14&   .111&.166&.075&.1\\
GR15&   .156& .121&.057 & .125\\
GR16&   .202& .076&.039&.15\\
GR17&   .248& .031& .021  &.175\\
GR18&   .278& 0& .008& .192\\ 

\hline

\end{tabular}\\

The range of reduction factors  is bounded
when the contribution from one of them becomes zero or greater than unity,
the extremes corresponding to no alignment and complete alignment
lying in the plane of the sky.
For ACAR, as $R_{g}$ increases from zero to a significant fraction,
the dominant changes are for the mantled
grains where $R_{gm}$ decreases from a substantial value to end the
range without alignment while $R_{sm}$ increases to substantial
alignment from smaller values, and  
the alignment efficiency of bare silicates reduces in compensation.
When graphite replaces ACAR the behaviour of the reduction factors
is broadly similar except that mantled silicates start the sequence with
$R_{sm} =0$.
Use of the MRN number ratio of carbon/silicate grains ($\simeq .9$)
gave a very much
truncated range of larger reduction factors than in the table.
As was found before by Lee and Draine (1985) mantled grains
need a greater alignment efficiency than bare to reproduce
the observations, and they concluded that the magnetic field must be highly
ordered and close to the plane of the sky, at least for the mantled grains.
The lower polarization efficiency for bare grains suggests a
poorer alignment mechanism of these grains compared with mantled grains
but this idea was not favoured by Lee and Draine, partly on account of the
lack of  excess polarization in GL 2591 (Dyck and Lonsdale 1981) even
though it shows strong ice absorption.

All of the sequences of polarization efficiencies in Table 1 yield
the linear polarization properties of BN in the 2.4--4$\mu$m region
to a good approximation
except for depolarization due to twist of the
alignment direction:  here the depolarization factor $D\gtsimeq 0.8$.

\begin{figure}
\includegraphics[width=8.5cm]{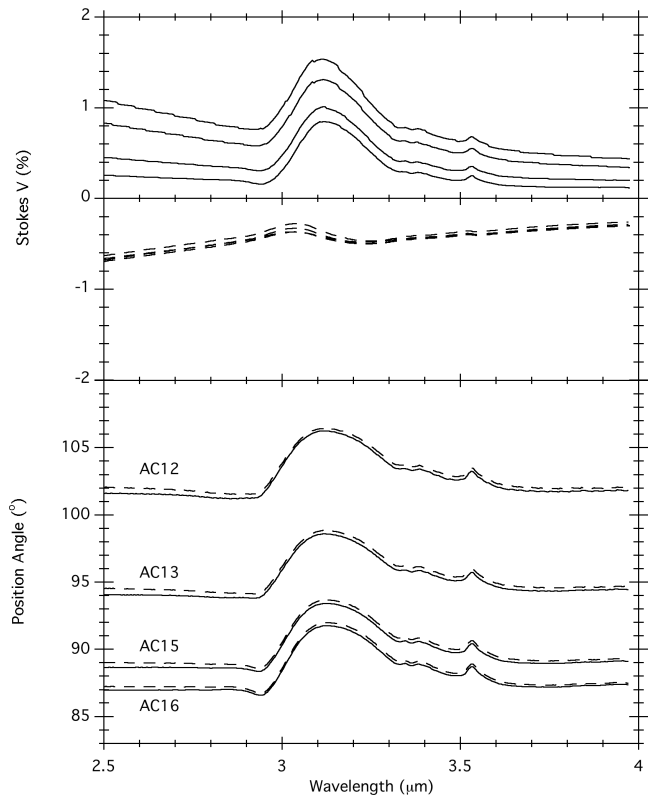}
\caption{Circular polarization (top) and position angle (bottom) for a sample of the two-slab sequences from Tables 1 and 2. Sequences in each panels increase from the top curve to the bottom, as shown. Reversal is shown by the dashed lines. The lower continuum with increasing sequence number is because the linear polarization continuum produced in slab1 is
decreasing at the expense of that produced in slab2. }
\label{fig4}
\end{figure}

\subsection{Simple two component models}
\subsubsection{A simple two-slab model}

To begin with a two slab model is considered in which only ice mantled
grains are present in the first slab and only bare grains in the second.

For a particular sequence in Table 1 the only
remaining free parameter is the angular twist, $\phi$,
between the slabs,
and this is varied until the position angle shift in the ice feature is
$\Delta\theta = +4^\circ$, which is that observed for BN.    
This value of $\Delta\theta$ is not attainable with sequence AC11
because when $R_g$ is near zero very little linear polarization
can be produced
in the second slab by the ``astrosilicate" alone, both $r_f$ and $r_c$
are small  and  even a twist
$\phi=50^{\circ}$ only produces $\Delta\theta = 1.5^{\circ}$
(cf equation 27, Martin 1974 and section 2).
When $R_{g}$= 0.025 (AC12) a twist between the slabs of 41$^{\circ}$ gives
the observed $\Delta\theta = +4^{\circ}$ and as $R_g$ increases the required
twist  reduces further.

Table 2 gives the derived circular polarizations from this simple 2-slab
model using the reduction factors from Table 1 and adjusting $\phi$
to give $\Delta\theta = +4^{\circ}$ when possible.
Purely for computing and display
convenience the two slab angles are equispaced about 90$^{\circ}$ to
give $\phi$, and 
$\theta_p$ is  the position angle at the peak of the ice
feature with reference to this frame.
The observed position angle of this feature is 118$^{\circ}$
for BN, and simple arithmetic enables the angles for slab1 to be found
in the BN frame and these are shown in Table 2
as $\theta1(BN)$.  In view of the large range of twist angle required
to give the observed position shift it is gratifying  that the
inferred icy slab angle is only 4.5-5$^{\circ}$ more positive
than the position angle in the feature, and remarkably independent
of the Table 1 sequence used.
A similar situation is evident for graphite
where  slab1 is just over 6$^{\circ}$ higher than the ice feature position
angle.
While $\theta$1 is well defined in the
two-slab model the same is not true for the bare grains where
$\theta2=\theta1 -\phi$ and is very dependent on the sequence from Table 1.

Fig 3 shows the flux, linear and circular polarization,
$P$ and $V$, and the position angle
for the sequence AC13, together with the effects of slab reversal,
shown dotted.  It
shows that there is a clear distinction between the
circular polarizations predicted when the icy region preceeds the bare
grains and the reverse.  In the former case the circular polarization is
similar in form and peak wavelength to the linear polarization and has
the same sign as the position angle shift; in the latter case the circular
polarization is reversed in sign and in place of the  feature
there is a weak `tilde' shape whose turning points positions differ
significantly from the single maximimum of the linear polarization.

Fig 4 shows the circular polarization and position angle spectra for the
different ACAR sequences for the range of $R_g$ in Table 1.
It shows that the
ice feature circular polarization changes little or not at all while
the continuum reduces in proportion to the value of $R_{gm}$.
The reason for this is that with increasing sequence number
the linear polarization continuum produced in slab1 
decreases at the expense of that produced in slab2. 
(Note that in both Fig 3 and Fig 4 the position angles are plotted in the
computing frame.)
In fig 4 the position angle shows very small shifts on
slab reversal, which is due to the full polarization transfer.
While the predicted position angles vary with the
reduction factors in this frame they remain at a fixed position from the alignment angle of slab1.

\begin{tabular}{cccccc}
\multicolumn{6}{c}{\bf Table 2}\\
\hline
\multicolumn{6}{c}{Derived Circular Polarization parameters} \\
\multicolumn{6}{c}{2-slab model} \\
\multicolumn{6}{c}{twist $\phi$ adjusted for $\Delta\theta=4^\circ$}\\
\hline
label &  degrees & \% & \% & degrees& degrees  \\
  & $\phi$ & $CP_p$ & $CP_c$ & $\theta_p$& $\theta$1(BN) \\
  \hline
\multicolumn{6}{c} {amorphous carbon (ACAR)}\\

\hline
AC11 &  -&-&-&-\\
AC12&   41&1.52&.787&106&122.5\\
AC13&   27&1.32&.62& 98.8&122.7\\
AC14&   22.8&1.23&.53&96.6&122.8\\
AC15&   16&.98&.34 & 93&123\\
AC16&   12.5&.82&.20& 91.5&122.75\\
AC17&   10.25&.69&.10& 90.7&122.4\\

\hline
 \multicolumn{6}{c} {graphite}\\

\hline
GR11 &  37&2.68&1.52&102&124.5\\
GR12&   33&2.62&1.44&99&125.5\\
GR13&   25&2.45&1.22&96&124.5\\
GR14&   21&2.32&.99&93.5&125\\
GR15&   17.6&2.12&.75 & 92&124.8\\
GR16&   15.2&1.95&.55&91&124.6\\
GR17&   13.2&1.80& .36& 90.5&124.1\\
GR18&   12& 1.67& .23& 90&124\\ 

\hline

\multicolumn{6}{l}{$CP_p$ and $CP_c$ is the circular polarisation at the peak }\\
\multicolumn{6}{l}{of the feature and in the nearby continuum, respectively.}\\

\end{tabular}

\subsubsection{A model with continuous twist}
With continuous twist the initial and final alignment angles
replace the
discrete slab angles and an extra free parameter allows us to
vary the twist in each region by choosing the position where the 
discrete change of material between the mantled and bare
grain regions occurs.
In Table 3 the entry under $\theta$1(BN) refers not to the extreme position
angle of slab1 but to the mean of slab1, and this too remains independent
of sequence number essentially the same as $\theta1$(BN) in the
two slab model.

Table 3 presents the results for the reduction factors of Table 1 for ACAR
applied to a model wherein the change between the two regions
occurs at the mid twist position, 90$^{\circ}$, and the regions
have equal twist.
As might
be expected the chief difference from Table 2 is that the angular
range of twist required is greater and also the circular polarizations
are larger.  AC15, for instance,  now
requires a twist of $\phi = 32^{\circ}$ and approximates the peak
and continuum circular polarizations of AC14 in the two-slab model. 
On reversal the `tilde' feature is slightly more prominent but still clearly
distinguishable from the unreversed case.

Keeping the total twist the same but varying the change angle
does not, perhaps surprisingly, produce  dramatic differences.
As most of the twist is transferred to the bare grain region
the circular polarization and the position angle shift decrease, while the
position angles themselves move closer to the start angle, but the changes
are not large and of order 10\% of the mid values.  As the
largest part of the twist is transfered to the icy region
all these changes assymptote to values only slightly
different from their mid-values. This is shown in Table 3 for AC15 where the
two components of twist in sections 1, 2 are shown; again $\theta1$(BN)
is unchanged.

\begin{tabular}{cccccc}
\multicolumn{6}{c}{\bf Table 3}\\
\hline
\multicolumn{6}{c}{Derived Circular Polarization parameters} \\
\multicolumn{6}{c}{continuous twist} \\
\multicolumn{6}{c}{twist $\phi$ adjusted for $\Delta\theta=4^\circ$}\\
\hline
label &  degrees & \% & \% & degrees& degrees  \\
  & $\phi$ & $CP_p$ & $CP_c$ & $\theta_p$ & $\theta$1(BN)\\
  \hline
\multicolumn{6}{c} {amorphous carbon (ACAR)}\\

\hline
AC11 &  -&-&-&-&-\\
AC12&   80&1.91&1.03&105.8&122.2\\
AC13&   54&1.66&.83&98.9&122.6\\
AC14&   45&1.52&.71& 96.6&122.6\\
AC15&   16,16&1.21&.48&93.3&122.7\\
"&  30,2&1.26&.46&86.1&122.9\\
" &  26,6&1.25&.47&88.2&122.8\\
"  & 6,26&1.14&.48&98.5&122.5\\
" &  2,30&1.10&.47&100.5&122.5\\
AC16&   24.5&.99&.31 & 91.6&122.4\\
AC17&   21&.84&.20& 90.7&122.5\\

\hline
 \multicolumn{6}{c} {graphite}\\

\hline
GR14&   41&2.68&1.27&93.4&124.6\\
GR16&   30&2.31&.83&91&124.5\\
GR17&   26&2.14& .64& 90.2&124.3\\

\hline

\end{tabular}\\

\subsubsection{An extension to the two component models}

In both the two-slab and continuous twist models the bare and mantled
grains have been strictly segregated and it is worth considering the
effect of mixing between the sections.  As mixing progresses the
position angle change, $\Delta\theta$, reduces rapidly while the
circular polarizations change little and to maintain the known
$\Delta\theta$ requires increasing the angular range, $\phi$. 

Alternatively, since it is well known that grains in the diffuse ISM
lack volatile mantles, mixing is likely to be confined to the presence
of a fraction of some of the bare grains in the mantled section.
As this fraction increases $\Delta\theta$ and both components of
circular polarization decrease so that the twist angle must be increased
to maintain  $\Delta\theta$ at its known value.
This imposes some limit on the dilution of the mantled region
by bare grains.  As an example AC13 in the simple two-slab model
requires a twist of 27$^{\circ}$ but if 20\% of the bare grains
are transferred to the mantled region then $\phi$ needs to be
increased to 35$^{\circ}$ to give the position angle shift
of +4$^{\circ}$.

\subsection{The alignment angles}
The large and model dependent twist angles between the sections,
especially if the twist is continous,  raise some uncertainty in
inferring alignment, and therefore magnetic field,
directions, $\theta_{1}, \theta_{2}$, in relation to the
observed position angle, $\theta$.
From Tables 2 and 3, however, $\theta1$(BN) is close to
the position angle of the ice feature (more positive by 4-4.5$^{\circ}$
for ACAR, 6-7$^{\circ}$  for graphite)
and independent of the unknown twist angle.  This is a property of the
linear polarization and only these relatively few 
and constant degrees separates the ice feature position angle
from the magnetic field direction in the icy grains.
The uncertainty in the alignment of the unmantled section
is much greater,
being different by the unknown twist from the first section.
However there
is a connection between the twist angle and the observable ratio of
circular polarization in the feature to that in the continuum and this
is seen in Fig 4:  a small ratio $CP_p/CP_c$ is associated with
a large twist, and vice versa.
Observations of circular polarization could
reduce some of these uncertainties.

\section{Conclusions}

The combination of linear and circular spectropolarimetry
can reveal the sequence of line of sight changes of grain chemistry
and magnetic field direction.  Frequently 
the H$_2$O ice feature is linearly polarized and has a different 
position angle to the continuum.  This implies not only fractionation
of material along the line of sight but also a twist of alignment,
and therefore field direction, and will be accompanied by circular
polarization.
Although the twist of alignment direction is model dependent and
can be very
many times greater than the observed position angle shift, the mean
alignment of the icy region is only 4-7$^{\circ}$ different from the
position angle of the ice feature, but it still leaves considerable
uncertainty in the alignment direction of the bare grains.

Circular polarimetry can clarify some of these uncertainties,
and in particular the ordering of material along the line of sight.
The presence of a circular polarization ice feature
accompanied by a position angle shift of the same sign (north through east)
means that the majority of the icy grains preceed
the majority of bare grains.  Circular polarization which lacks
the ice feature and is of opposite sign to the position angle shift
means the majority of icy grains overlie the majority of bare grains.

Circular spectropolarimetry of BN has been obtained under very
poor conditions and over an inadequate wavelength range of 2.9-3.3$\mu$m
which does not sample the continuum: 
the spectrum is noisy without clear feature but is clearly positive,
also  shown by Serkowski and Rieke (1973)at 3.45$\mu$m.
This, together with the sign of the position angle change,
is one of the conditions  that the icy grains along the
line of sight to BN are overlaid by bare grains.
Two other sources also show the correlation of the position
angle shift and the circular polarization.  These are GL490 and GL2591
which Lonsdale etal (1980) find to have circular polarizations at 2.2$\mu$m
of -0.4\% and -.85\% respectively and have position angle shifts of
-4$^{\circ}$ and -1$^{\circ}$, also respectively (Holloway etal 2002).  In these too the icy
grains are overlaid by bare grains.
The other condition is the signature of the ice-feature itself in the
circular polarization.   
It is important that good quality linear and
circular spectropolarimetry is
obtained and extended to other sources including
other regions of the BNKL complex.  
The form of the circular polarization spectrum also contains additional
information related to the 
chemistries of the different regions and contrains the
twist angle between the regions.

{\bf Acknowledgements}

We thank the referee, Patrick Roche, for useful comments and contribution.

\appendix

\section{Polarization Transfer Equations}

The constraints of extinction and polarization in section 3 lead
to the column
densities of the constituents as well as the range of reduction factors in
Table 1. These are applied to the individual $C_{abs}, C_{pol}$ and
$C_{crc}$ of the constituents to produce a combined $C$ for each slab or
section.  Starting from unpolarized flux of unit intensity $I$
the transfer
equations below are applied and iterated  through the column density of the
first slab.  The resultant $I,Q,U,V$ are then used as input to the
second slab/section and integrated through its column density. 
The process is reversed to produce the
output when section 2 preceeds section1.

In the following $C_{abs}$, $C_{pol}$, and $C_{crt}$ refer to the combined
values incuding the relevant $R$'s and of $N$'s (column densities)
of the slab constituents.

$$
\frac{dI}{dz} = \frac{-C_{abs}}{2}I +
    \frac{C_{pol}}{2}(Q\,\textrm{cos}\,2\theta + U\,\textrm{sin}\,2\theta) 
$$

$$
\frac{dQ}{dz} = \frac{-C_{abs}}{2}Q + \frac{C_{pol}}{2}\,I\,\textrm{cos}\,2\theta + C_{pha}\,V\,\textrm{sin}\,2\theta  
$$

$$
\frac{dU}{dz} = \frac{-C_{abs}}{2}U + \frac{C_{pol}}{2}\,I\,\textrm{sin}\,\theta - C_{pha}\,V\,\textrm{cos}\,2\theta  
$$

$$
\frac{dV}{dz} = \frac{-C_{abs}}{2}V + C_{pha}(U\textrm{cos}\,2\theta
     - Q\,\textrm{sin}2\,\theta) 
$$

In the case of the continuous twist models $\theta$ is incremented through the
integration.

\end{document}